\documentclass[12pt]{article}
\usepackage{amsfonts}
\begin{document}
\date{}

\title{\textbf{Gauge Generators, Transformations and Identities on a Noncommutative Space}}
\author{{Rabin Banerjee}\thanks{E-mail: rabin@bose.res.in} \ and {Saurav Samanta}\thanks{E-mail: saurav@bose.res.in}\\
\\\textit{S.~N.~Bose National Centre for Basic Sciences,}
\\\textit{JD Block, Sector III, Salt Lake, Kolkata-700098, India}}
\maketitle
                                                                                
\begin{quotation}
\noindent \normalsize 
By abstracting a connection between gauge symmetry and gauge identity on a noncommutative space, we analyse star (deformed) gauge transformations with usual Leibniz rule as well as undeformed gauge transformations with a twisted Leibniz rule. Explicit structures of the gauge generators in either case are computed. It is shown that, in the former case, the relation mapping the generator with the gauge identity is a star deformation of the commutative space result. In the latter case, on the other hand, this relation gets twisted to yield the desired map.
\end{quotation}
\section{Introduction}
Recent analysis\cite{wess,jwess,vas,chai,gaume,wes} of gauge transformations in noncommutative theories reveals that, in extending gauge symmetries to the noncommutative space-time, there are two distinct possibilities. Gauge transformations are either deformed such that the standard comultiplication (Leibniz) rule holds or one retains the unmodified gauge transformations as in the commutative case at the expense of altering the Leibniz rule. In the latter case the new rule to compute the gauge variation of the star products of fields results from a twisted Hopf algebra of the universal enveloping algebra of the Lie algebra of the gauge group extended by translations.

While both these approaches preserve gauge invariance of the action, there is an important distinction. In the case of deforming gauge transformations into star gauge transformations, gauge symmetries act only on the fields in a similar way as in theories on commutative space time. Star gauge symmetry can thus be interpreted as a physical symmetry in the usual sense. On the contrary if ordinary gauge transformations are retained and a twisted Leibniz rule is implemented, then the transformations do not act only on the fields. Consequently it is not a physical symmetry in the conventional sense and its connection with the previous case also remains obscure\cite{gaume}.

In this paper we analyse both these approaches within a common framework which also illuminates a correspondence with the treatment of gauge symmetry in commutative space time. To do this we remind that there is a general method of discussing gauge symmetries, either in the Lagrangian or Hamiltonian formulations, for theories on commutative space-time. We shall here concentrate on the Lagrangian version\cite{gitman}. It is known that, corresponding to each gauge symmetry, there is a gauge identity that is expressed in terms of the Euler derivatives. Moreover, this identity also involves the generator of infinitesimal gauge transformations in a very specific manner.

We extend this analysis to noncommutative gauge theories. A relation between the gauge generator
and the gauge identity is derived. It is found to be a star deformation of the relation found in the usual commutative picture. From this relation and a knowledge of the gauge identity (obtained by a simple inspection) the explicit form of 
the gauge generator is derived. Then the other viewpoint of keeping the gauge transformation undeformed at the price of a twisted Leibniz rule is considered. The generator of the undeformed gauge transformation is derived.  Its structure is
shown to be similar to the commutative space expression.  Furthermore, we find that the relation connecting
the gauge  generator with the gauge identity (which is form invariant irrespective of whether star or twisted gauge transformations
are being considered) is neither the undeformed result nor its star deformation, as obtained in the previous treatment. Rather, it is a twisted form of the conventional (undeformed) result.

The paper is organised as follows. In section 2 we briefly review the situation where both interpretations of noncommutative gauge symmetry lead to identical conservation laws. Sections 3 and 4 give a detailed account of the computations for deformed gauge symmetry with standard Leibniz rule and undeformed gauge symmetry with twisted Leibniz rule, respectively. Explicit expressions for the generators and their connection with the gauge identity is analysed. Section 5 is a summary.  
\section{Gauge Transformations and Conservation Law}
Consider a theory on noncommutative space time whose dynamics is governed by the action{\footnote{From the beginning we are considering a non-Abelian theory.}},
\begin{eqnarray}
S=\int \textrm{d}^4x \ [-\frac{1}{2}\textrm{Tr}(F_{\mu\nu}(x)*F^{\mu\nu}(x))+\bar{\psi}(x)*(i\gamma^{\mu}D_{\mu}*-m)\psi(x)]
\label{lag}
\end{eqnarray}
where
\begin{eqnarray}
&&D_{\mu}*\psi(x)\equiv \partial_{\mu}\psi(x)+igA_{\mu}(x)*\psi(x)\\
&&F_{\mu\nu}(x)\equiv\partial_{\mu}A_{\nu}(x)-\partial_{\nu}A_{\mu}(x)+ig[A_{\mu}(x),A_{\nu}(x)]_*
\label{f}
\end{eqnarray}
Here the star commutator is given by
\begin{eqnarray}
[A_{\mu}(x),A_{\nu}(x)]_*=A_{\mu}(x)*A_{\nu}(x)-A_{\nu}(x)*A_{\mu}(x)
\end{eqnarray}
while the star product is defined as usual
\begin{eqnarray}
(f*g)(x)={\textrm{exp}}\left(\frac{i}{2}\theta^{\mu\nu}\partial_{\mu}^x\partial_{\nu}^y\right)f(x)g(y)|_{x=y}
\label{str}
\end{eqnarray}
where $\theta^{\mu\nu}$ is a constant two index antisymmetric object.

The above action describes the noncommutative version of a non-Abelian theory which includes both the gauge and a matter (fermionic) sector with a proper interaction term. This action is invariant under both deformed gauge transformations,
\begin{eqnarray}
\begin{array}{rcl}
&&\delta_* A_{\mu}=\mathcal{D}_{\mu}*\eta=\partial_{\mu}\eta+ig(A_{\mu}*\eta-\eta*A_{\mu}),\\
&&\delta_* F_{\mu\nu}=ig[F_{\mu\nu},\eta]_*=ig(F_{\mu\nu}*\eta-\eta*F_{\mu\nu})\\
&&\delta_* \psi=-ig\eta*\psi\\
&&\delta_* \bar{\psi}=ig\bar{\psi}*\eta
\end{array}
\label{XX}
\end{eqnarray}
with the usual Leibniz rule,
\begin{eqnarray}
\delta_* (f*g)=(\delta_* f)*g+f*(\delta_* g)
\label{tX}
\end{eqnarray}
as well as the undeformed gauge transformations
\begin{eqnarray}
\begin{array}{rcl}
&&\delta_{\eta} A_{\mu}=\mathcal{D}_{\mu}\eta=\partial_{\mu}\eta+ig(A_{\mu}\eta-\eta A_{\mu}),\\
&&\delta_{\eta} F_{\mu\nu}=ig[F_{\mu\nu},\eta]=ig(F_{\mu\nu}\eta-\eta F_{\mu\nu})\\
&&\delta_{\eta} \psi=-ig\eta\psi\\
&&\delta_{\eta} \bar{\psi}=ig\bar{\psi}\eta
\end{array}
\label{YY}
\end{eqnarray}
with the twisted Leibniz rule\cite{wess,jwess,vas},
\begin{eqnarray}
\delta_{\eta}(f*g)&=&\sum_n(\frac{-i}{2})^n\frac{\theta^{\mu_1\nu_1}\cdot \cdot \cdot\theta^{\mu_n\nu_n}}{n!}\nonumber\\
&&(\delta_{\partial_{\mu_1}\cdot \cdot \cdot\partial_{\mu_n}\eta}f*\partial_{\nu_1}\cdot \cdot \cdot\partial_{\nu_n}g+\partial_{\mu_1}\cdot \cdot \cdot\partial_{\mu_n}f*\delta_{\partial_{\nu_1}\cdot \cdot \cdot\partial_{\nu_n}\eta}g)
\label{co}
\end{eqnarray}
The essence of this modified rule is that the gauge parameter $\eta$ always remains outside the star operation
and hence is unaffected by it. This fact becomes important when we discuss the twisted gauge symmetry (\ref{YY},\ref{co}).

It is obvious from (\ref{f}) and the definition of the gauge transformations (\ref{XX}) that, in general, both $A_{\mu}$ as well as $F_{\mu\nu}$ are enveloping algebra valued for deformed gauge symmetry. For the case of twisted gauge symmetry (\ref{YY},\ref{co}), however, one has to consider the equation of motion derived later (see (\ref{eqn})), interpreted as equations for the gauge field $A_{\mu}$, to conclude that here also $A_{\mu}$ is enveloping algebra valued. The field tensor $F_{\mu\nu}$, by its very definition (\ref{f}), is of course enveloping algebra valued. Thus, in both treatments of gauge symmetry, $A_{\mu}$ and $F_{\mu\nu}$ are enveloping algebra valued\cite{jwess}. This implies that the gauge potential $A_{\mu}$ has to be expanded over a basis of the vector space spanned by the homogeneous polynomials in the generators of the Lie algebra,
\begin{eqnarray}
A^{\mu}(x)&=&A^{\mu}_a(x):T^a:+A^{\mu}_{a_1a_2}(x):T^{a_1}T^{a_2}:\nonumber\\&&+...A^{\mu}_{a_1a_2...a_n}(x):T^{a_1}T^{a_2}...T^{a_n}:+...
\label{amu}
\end{eqnarray}
where the double dots indicate totally symmetrised products,
\begin{eqnarray}
:T^a:&=&T^a\nonumber\\
:T^{a_1}T^{a_2}:&=&\frac{1}{2}\{T^{a_1},T^{a_2}\}=\frac{1}{2}(T^{a_1}T^{a_2}+T^{a_2}T^{a_1})
\label{ta}\\
:T^{a_1}...T^{a_n}:&=&\frac{1}{n!}\sum_{\pi\in S_n}T^{a_{\pi(1)}}...T^{a_{\pi(n)}}\nonumber
\end{eqnarray}
These symmetrised products may be simplified by using the basic Lie algebraic relation,
\begin{eqnarray}
&&[T^a,T^b]=if^{abc}T^c
\label{fabc}
\end{eqnarray}
where $f^{abc}$ are the structure constants.

To be specific let us take the case of $SU(2)$ in the two dimensional representation. In this representation the generators are the $2\times 2$ Pauli matrices ($T^a=\frac{\sigma^a}{2}$) satisfying (\ref{fabc}) with $f^{abc}=\epsilon^{abc}$ and,
\begin{eqnarray}
\{\frac{\sigma^a}{2},\frac{\sigma^b}{2}\}=\frac{1}{2}\delta^{ab} \  \  \  \ (a=1,2,3).
\end{eqnarray}
We may thus write $A_{\mu}$ as follows:
\begin{eqnarray}
A_{\mu}=B_{\mu}+A_{\mu}^a\sigma^a \  \  \  \ (a=1,2,3).
\end{eqnarray}
It is also possible to interpret this situation as giving rise to the standard representation of $U(2)$ with its four generators $(\mathbb{I},\sigma^a)$
\begin{eqnarray}
A_{\mu}=A_{\mu}^aT^a \  \  \  (T^a=\mathbb{I},\sigma^a)
\end{eqnarray}
Similarly in the three dimensional representation of $SU(2)$, the generators are defined in the adjoint representation $(T^a)^{bc}=-i\epsilon^{abc}$. Now the process of symmetrisation (\ref{ta}) yields nine $3\times 3$ linearly independent Hermitian matrices (these are the three $T^a$'s and the six $\{T^a,T^b\}$) and hence we obtain the standard representation of $U(3)$. This implies that the enveloping algebra valued $A_{\mu}$ given in (\ref{amu}) is equivalently simplified to a Lie algebraic $U(3)$ representation with $A_{\mu}=A_{\mu}^aT^a$, where $T^a$'s are the nine generators of $U(3)$.

In general one can verify that the generators given by (\ref{ta}), apart from forming a Lie algebra (\ref{fabc}), also close under anticommutation\cite{amorim,rabin},
\begin{eqnarray}
&&\{T^a,T^b\}=d^{abc}T^c.
\label{dabc}
\end{eqnarray}
The simpler nontrivial algebra that matches these conditions is $U(N)$ in the representation given by $N\times N$ hermitian matrices.

Following \cite{banora} it is feasible to choose $T^1=\frac{1}{\sqrt{2N}}\mathbb{I}_{(N\times N)}$ and the remaining $N^2-1$ of the $T$'s as in $SU(N)$. Then the trace condition also follows as,
\begin{eqnarray}
{\textrm{Tr}}(T^aT^b)=\frac{1}{2}\delta^{ab}
\label{trace}
\end{eqnarray}
and $f^{abc}$, $d^{abc}$ are completely antisymmetric and completely symmetric respectively.

In the rest of this paper we will work with these simplifications. The gauge potential and the field strength will be explicitly written as,
\begin{eqnarray}
A_{\mu}&=&A_{\mu}^aT^a\\
F_{\mu\nu}&=&F_{\mu\nu}^aT^a
\end{eqnarray}
where the $T^a$'s are the $N^2$ hermitian generators of $U(N)$ that satisfy the condition (\ref{fabc}), (\ref{dabc}) and (\ref{trace}).

In order to derive the field equations we need the following well known properties of the star product within an integral
\begin{eqnarray}
\int \textrm{d}x \ A(x)*B(x)=\int \textrm{d}x \ A(x)B(x)=\int \textrm{d}x \ B(x)*A(x)
\label{b1}
\end{eqnarray}
and a special case of which is given by,
\begin{eqnarray}
\int \ (A*B*C)=\int \ (B*C*A)=\int \ (C*A*B)
\label{b2}
\end{eqnarray}
First, we vary the gauge field $A$ to see the variation of the action (\ref{lag}). We follow the convention of keeping the field to be varied at the extreme left. Then we get,
\begin{eqnarray}
\frac{\delta S}{\delta A_{\sigma}(y)}&=&\int d^4x \ [-\textrm{Tr} \ \frac{\delta F_{\mu\nu}(x)}{\delta A_{\sigma}(y)}*F^{\mu\nu}(x)+\frac{\delta}{\delta A_{\sigma}(y)}(-g\bar{\psi}*\gamma^{\nu}A_{\nu}*\psi)]\nonumber
\end{eqnarray}
where, in the first integral, (\ref{b1}) has been used. Further simplifications are done by using the cyclicity property (\ref{b2}) in the second integral of the above expression. We obtain,
\begin{eqnarray}
\frac{\delta S}{\delta A_{\sigma}(y)}&=&\int d^4x \ [-\textrm{Tr} \ \frac{\delta}{\delta A_{\sigma}(y)}(\partial_{\mu}A_{\nu}-\partial_{\nu}A_{\mu}-i[A_{\mu},A_{\nu}]_*)*F^{\mu\nu}(x)\nonumber\\
&&+\frac{\delta}{\delta A_{\sigma}(y)}(gA_{\nu}*\psi_j\gamma^{\nu}_{ij}*\bar{\psi}_i)]\nonumber
\end{eqnarray}
The extra negative sign in the second integral is due to the flip of two grassmanian variables. Finally using the fact that $F^{\mu\nu}$ is antisymmetric in its indices, we obtain
\begin{eqnarray}
\frac{\delta S}{\delta A_{\sigma}(y)}&=&\int d^4x \ [-2\textrm{Tr} \ \frac{\delta}{\delta A_{\sigma}(y)}(\partial_{\mu}A_{\nu}-iA_{\mu}*A_{\nu})*F^{\mu\nu}(x)\nonumber\\
&&+\frac{\delta A_{\nu}(x)}{\delta A_{\sigma}(y)}*(g\psi_j\gamma^{\nu}_{ij}*\bar{\psi}_i)]\nonumber\\
&=&\int d^4x \ [2\textrm{Tr} \ \frac{\delta A_{\nu}(x)}{\delta A_{\sigma}(y)}*(\partial_{\mu}F^{\mu\nu}(x)-iA_{\mu}*F^{\mu\nu}+iF^{\mu\nu}*A_{\mu})\nonumber\\
&&+\frac{\delta A_{\nu}(x)}{\delta A_{\sigma}(y)}*(g\psi_j\gamma^{\nu}_{ij}*\bar{\psi}_i)].
\end{eqnarray}
The invariance of the action together with (\ref{trace}) now leads to the equation of motion for the gauge field,
\begin{eqnarray}
\partial_{\mu}F^{\mu\nu}+ig[A_{\mu},F^{\mu\nu}]_*+j^{\nu}=0
\label{eqn}
\end{eqnarray}
where $j^{\nu}$ is the fermionic current
\begin{eqnarray}
j^{\nu}=g\psi_j(\gamma^{\nu})_{ij}*\bar{\psi}_i.
\label{14a1}
\end{eqnarray}
The variation of the matter field $\bar{\psi}$ in the action (\ref{lag}) yields
\begin{eqnarray}
\frac{\delta S}{\delta \bar{\psi}(y)}&=&\int d^4x \ \frac{\delta}{\delta \bar{\psi}(y)}(\bar{\psi}*i\gamma^{\mu}\partial_{\mu}\psi-g\bar{\psi}*\gamma^{\mu}A_{\mu}*\psi-m\bar{\psi}*\psi)\nonumber\\
&=&\int d^4x \ \frac{\delta \bar{\psi}(x)}{\delta \bar{\psi}(y)}*(i\gamma^{\mu}\partial_{\mu}\psi-g\gamma^{\mu}A_{\mu}*\psi-m\psi)
\end{eqnarray}
which gives the equation of motion
\begin{eqnarray}
i\gamma^{\mu}\partial_{\mu}\psi-g\gamma^{\mu}A_{\mu}*\psi-m\psi=0.
\label{eqpsi}
\end{eqnarray}
Similarly for the other matter field we get the equation of motion
\begin{eqnarray}
i\partial_{\mu}\bar{\psi}\gamma^{\mu}+g\bar{\psi}*\gamma^{\mu}A_{\mu}+m\bar{\psi}=0.
\label{eqpsibar}
\end{eqnarray}
Operating $\partial_{\nu}$ on eq. (\ref{eqn}) we get a current conservation law\cite{wes}
\begin{eqnarray}
\partial_{\nu}J^{\nu}=0; \ J^{\nu}\equiv ig[A_{\mu},F^{\mu\nu}]_*+j^{\nu}
\label{j}
\end{eqnarray}
This can be explicitly verified from the definition of $J^{\nu}$ as follows:
\begin{eqnarray}
\partial_{\nu}J^{\nu}&=&ig[\partial_{\nu}A_{\mu},F^{\mu\nu}]_*+ig[A_{\mu},\partial_{\nu}F^{\mu\nu}]_*+\partial_{\nu}j^{\nu}\nonumber\\
&=&-\frac{1}{2}ig[\partial_{\mu}A_{\nu}-\partial_{\nu}A_{\mu},F^{\mu\nu}]_*+ig[A_{\mu},ig[A_{\nu},F^{\nu\mu}]_*+j^{\mu}]_*+\partial_{\nu}j^{\nu}\nonumber\\
&=&-\frac{1}{2}ig[F_{\mu\nu},F^{\mu\nu}]_*+\frac{(ig)^2}{2}[[A_{\mu},A_{\nu}]_*,F^{\mu\nu}]_*-(ig)^2[A_{\mu},[A_{\nu},F^{\mu\nu}]_*]_*\nonumber\\&&+ig[A_{\mu},j^{\mu}]_*+\partial_{\nu}j^{\nu}.\nonumber
\end{eqnarray}
Since the first term of the right hand side vanishes trivially we write the above expression as
\begin{eqnarray}
\partial_{\nu}J^{\nu}&=&\frac{(ig)^2}{2}\left([[A_{\mu},A_{\nu}]_*,F^{\mu\nu}]_*-[A_{\mu},[A_{\nu},F^{\mu\nu}]_*]_*+[A_{\nu},[A_{\mu},F^{\mu\nu}]_*]_*\right)\nonumber\\&&+ig[A_{\mu},j^{\mu}]_*+\partial_{\nu}j^{\nu}.\nonumber
\end{eqnarray}
The term in the parentheses vanishes from the Jacobi identity and we obtain,
\begin{eqnarray}
\partial_{\nu}J^{\nu}=ig[A_{\mu},j^{\mu}]_*+\partial_{\nu}j^{\nu}.
\end{eqnarray}
(Star) multiplying eq. (\ref{eqpsi}) by $-ig\bar{\psi}$ from right and eq. (\ref{eqpsibar}) by $-ig\psi$ from left and then adding those two equations we find,
\begin{eqnarray}
\partial_{\nu}J^{\nu}=ig[A_{\mu},j^{\mu}]_*+\partial_{\nu}j^{\nu}=0
\end{eqnarray}
where the definition (\ref{14a1}) of $j^{\mu}$ has been used. It is also possible to obtain the current defined in eq. (\ref{j}) from (\ref{lag}) by using a Noether-like procedure\cite{gaume}. If we make the following ``global" transformation on the gauge and matter fields,
\begin{eqnarray} 
&&\delta A_{\mu}(x)=ig[\omega(x),A_{\mu}(x)]_*\\
&&\delta\psi(x)=-ig\omega(x)*\psi(x)\\
&&\delta\bar{\psi}(x)=ig\bar{\psi}(x)*\omega(x)
\end{eqnarray}
and set $\omega(x)$ to a constant at the end of the calculation, the conserved current (\ref{j}) follows from (\ref{lag}).

As has been stressed\cite{gaume} the conservation law (\ref{j}) is compatible with both types of gauge symmetry (\ref{XX}) (with the Leibniz rule (\ref{tX})) and (\ref{YY}) (with the Leibniz rule (\ref{co})). One finds for instance, $\partial_{\mu}(\delta J^{\mu})=\partial_{\mu}(\delta_{\eta} J^{\mu})=0$. It is clear that the conservation law is unable to provide any distinction between the two types of gauge transformations. This is not surprising since this conservation law is an on shell symmetry which is quite distinct from gauge symmetry which is an off-shell symmetry. So in the next two sections we study the gauge (both star gauge and twisted gauge) symmetry of the system where on shell considerations are discarded.
The difference between the star gauge and twisted gauge symmetries thereby gets manifested.

\section{Analysis for Star gauge transformation}
As already stated, the presence of gauge symmetry is characterised by an identity which is called the gauge identity. In this section we first discuss a general formalism to connect this identity with the gauge generator that eventually leads to the gauge transformations. Next we use this method for the particular model (\ref{lag}) to find the gauge generators from which the star deformed gauge transformations (\ref{XX}) are systematically obtained.

Consider a general form of the action on noncommutative space as\footnote{Here we adopt the notation $x$ for the four vector $x^{\mu}=({\bf{x}},t)$.},
\begin{eqnarray}
S=\int\textrm{d}t \ L=\int \textrm{d}^4x \  \mathcal{L}
\left(q_{\alpha}({\bf{x}},t), \ \partial_iq_{\alpha}({\bf{x}},t), \ \partial_tq_{\alpha}({\bf{x}},t)\right)
\label{L}
\end{eqnarray}
where $\alpha$ denotes the number of fields. An arbitrary variation of this action can be written as
\begin{eqnarray}
\delta S=-\int \textrm{d}^4x \ \delta q^{\alpha}({\bf{x}},t)*L_{\alpha}({\bf{x}},t)
\label{l}
\end{eqnarray}
where the vanishing of the Euler derivative $L$ yields the equations of motion. Now if we vary the field $q^{\alpha}$ in the following way
\begin{eqnarray}
\delta q^{\alpha}({\bf{x}},t)=\sum_{s=0}^n(-1)^s\int\textrm{d}^3{\bf{z}} \ \frac{\partial^s\eta^b({\bf{z}},t)}{\partial t^s}*\rho^{\alpha b}_{(s)}(x,z)
\label{a}
\end{eqnarray}
with $\eta$ and $\rho$ being the parameter and generator, respectively, of the transformation, the variation of the action can be written from (\ref{l}) as
\begin{eqnarray}
\delta S&=&-\int\textrm{d}^4x \ \int\textrm{d}^3{\bf{z}} \ \eta^b({\bf{z}},t)*\rho^{\alpha b}_{(0)}(x,z)*L_{\alpha}({\bf{x}},t)-\nonumber\\
&&\int \textrm{d}^4x \  \sum_{s=1}^n(-1)^s\int \textrm{d}^3{\bf{z}} \ \frac{\partial}{\partial t}\left(\frac{\partial^{s-1}\eta^b({\bf{z}},t)}{\partial t^{s-1}}\right)*\rho^{\alpha b}_{(s)}(x,z)*L_{\alpha}({\bf{x}},t)\nonumber\\
&=&-\int\textrm{d}^4x \ \int\textrm{d}^3{\bf{z}} \ \eta^b({\bf{z}},t)*\rho^{\alpha b}_{(0)}(x,z)*L_{\alpha}({\bf{x}},t)-\nonumber\\
&&\int \textrm{d}^4x \sum_{s=1}^n(-1)^{s-1}\int \textrm{d}^3{\bf{z}} \ \frac{\partial^{s-1}\eta^b({\bf{z}},t)}{\partial t^{s-1}}*\frac{\partial}{\partial t}\left(\rho^{\alpha b}_{(s)}(x,z)*L_{\alpha}({\bf{x}},t)\right)\nonumber\\
&=&-\int \textrm{d}^4z \ \eta^b({\bf{z}},t)*\left(\int \textrm{d}^3{\bf{x}} \ \rho^{\alpha b}_{(0)}(x,z)*L_{\alpha}({\bf{x}},t)\right)-\nonumber\\
&&\int \textrm{d}^4z \ \eta^b({\bf{z}},t)*\left(\int \textrm{d}^3{\bf{x}} \ \frac{\partial}{\partial t}(\rho^{\alpha b}_{(1)}(x,z)*L_{\alpha}({\bf{x}},t))\right)-\cdot\cdot\cdot
\label{mann}
\end{eqnarray}
Eq. (\ref{mann}) is written in the compact form
\begin{eqnarray}
\delta S=-\int\textrm{d}^4z \ \eta^a({\bf{z}},t)*\Lambda^a({\bf{z}},t)
\label{bak}
\end{eqnarray} 
where\footnote{Equations (\ref{bak}) and (\ref{lam}) are the star deformed version of the commutative space results given, for instance, in \cite{gitman,shirzad}.}
\begin{eqnarray}
\Lambda^a({\bf{z}},t)=\left[\sum_{s=0}^n\int \textrm{d}^3{\bf{x}} \ \frac{\partial^s}{\partial t^s}\left(\rho^{\alpha a}_{(s)}(x,z)*L_{\alpha}({\bf{x}},t)\right)\right].
\label{lam}
\end{eqnarray}
If the action is invariant ($\delta S=0$), then it implies,
\begin{eqnarray}
\Lambda^a({\bf{z}},t)=0.
\end{eqnarray}

The last equality must be identically valid without use of any equation of motion. It is called the gauge identity. Eq. (\ref{a}) defines the gauge transformation of the fields with $\rho$ being the generator. Furthermore, the gauge identity involves the generator and Euler derivatives in a specific fashion given by (\ref{lam}).

There are now two ways to apply this general formulation to a specific gauge model. Starting from a knowledge of the gauge transformations $\delta q$ (obtained, for instance, by inspection) it should be possible to compute the generators $\rho$ by using (\ref{a}). Then the explicit structure for the gauge identity follows from (\ref{lam}). Alternatively, one starts from the gauge identity (obtained, as shown here later on), inverts the process, finally generating the gauge transformations. Here we adopt the second approach for the model (\ref{lag})

 The first step to obtain the gauge identity is to derive the Euler derivatives. This is simply done by considering an arbitrary variation of the action (\ref{lag}), expressed in terms of the variations of the basic fields,
\begin{eqnarray}
\delta S=-\int \textrm{d}^4x \ \delta A_{\mu}^{a}*L^{\mu a}+\delta\psi_i*L_i+\delta\bar{\psi}_i*L'_i
\label{tm}
\end{eqnarray}
where the Euler derivatives $L_{\mu}^a, \ L_i$ and $L'_i$ are given by
\begin{eqnarray}
&&L^{\mu a}=-\left(\mathcal{D}_{\sigma}*F^{\sigma\mu}\right)^a-g\psi_j(\gamma^{\mu}T^a)_{ij}*\bar{\psi}_i
\label{eu1}\\
&&L_i=-i\partial_{\mu}\bar{\psi}_j(\gamma^{\mu})_{ji}-g\bar{\psi}_j*(\gamma^{\mu}A_{\mu}^aT^a)_{ji}-m\bar{\psi}_i
\label{eu2}\\
&&L_i'=-i(\gamma^{\mu})_{ij}\partial_{\mu}{\psi}_j+g(\gamma^{\mu}A_{\mu}^aT^a)_{ij}*{\psi}_j+m\psi_i.
\label{eu3}
\end{eqnarray}
Here the covariant derivative $\mathcal{D}$ is defined in the adjoint representation,
\begin{eqnarray}
&&\mathcal{D}_{\mu}*\xi=\partial_{\mu}\xi+ig[A_{\mu},\xi]_*;\\
&&(\mathcal{D}_{\mu}*\xi)^a=\partial_{\mu}\xi^{a}-\frac{g}{2}f^{abc}\{A^b_{\mu},\xi^{c}\}_*+i\frac{g}{2}d^{abc}[A^b_{\mu},\xi^{c}]_*
\label{D}
\end{eqnarray}
where we have used (\ref{fabc}) and (\ref{dabc}). We now define a quantity $\Lambda$, involving the various Euler derivatives of the system as,
\begin{eqnarray}
\Lambda^a=-\left(\mathcal{D}^{\mu}*L_{\mu}\right)^a-igT^a_{ij}{\psi}_j*L_i-igT^a_{ji}L_i'*\bar{\psi}_j.
\label{lambdaa}
\end{eqnarray}
Exploiting the definitions of the covariant derivative (\ref{D}) and Euler derivatives (\ref{eu1},\ref{eu2},\ref{eu3}) the above expression, by an explicit calculation, is found out to be zero, i. e. it vanishes identically without using any equations of motion,
\begin{eqnarray}
\Lambda^a=-\left(\mathcal{D}^{\mu}*L_{\mu}\right)^a-igT^a_{ij}{\psi}_j*L_i-igT^a_{ji}L_i'*\bar{\psi}_j=0.
\label{lambda}
\end{eqnarray}

The above relation is the cherished gauge identity for the model (\ref{lag}). It is important to note that the structure of $\Lambda^a$ in (\ref{lambdaa}) is similar to the general form (\ref{lam}) in the sense that it involves the appropriate Euler derivatives. By a comparison of the two, the generators $\rho$ are obtained. To this end let us now write (\ref{lam}) in a convenient way which is more suitable for our particular model,
\begin{eqnarray}
\Lambda^a({\bf{z}},t)&=&\sum_s\int \textrm{d}^3{\bf{x}} \ \frac{\partial^s}{\partial t^s}\left(\rho^{b\mu a}_{(s)}(x,z)*L^b_{\mu}({\bf{x}},t)\right)+\nonumber\\
&&\sum_s\int \textrm{d}^3{\bf{x}} \ \frac{\partial^s}{\partial t^s}\left(\phi^a_i(x,z)*L_i({\bf{x}},t)+\phi'^a_i(x,z)*L'_i({\bf{x}},t)\right).
\label{eq}
\end{eqnarray}
The values of the generators $\rho$, $\phi$ and $\phi'$ are obtained by comparing eqs. (\ref{lambdaa}) and (\ref{eq}). Since the calculations involve some subtlety due to the noncommutative nature of the coordinates, couple of intermediate steps are presented here. The contribution coming from the zeroth component of the gauge field Euler derivative $L_{\mu}$ can be written from (\ref{lambda}) as
\begin{eqnarray}
\Lambda^a|_{L_0}&=&-\left(\mathcal{D}^{0}*L_{0}\right)^a\nonumber\\
&=&-\frac{g}{2}f^{abc}\{L_0^b,A^{0c}\}_*+i\frac{g}{2}d^{abc}[L_0^b,A^{0c}]_*-\frac{\partial}{\partial t}L_{0}^a.
\end{eqnarray}
We write the above equation in the following form
\begin{eqnarray}
\Lambda^a|_{L_0}({\bf{z}},t)&=&-\frac{g}{2}f^{abc}\int \textrm{d}^3{\bf{x}} \ \left(L_0^b(x)*A^{0c}(x)+A^{0c}(x)*L_0^b(x)\right)*\delta^3({\bf{x}}-{\bf{z}})\nonumber\\
&&-\frac{i}{2}d^{abc}\int \textrm{d}^3{\bf{x}} \ \left(L_0^b(x)*A^{0c}(x)-A^{0c}(x)*L_0^b(x)\right)*\delta^3({\bf{x}}-{\bf{z}})\nonumber\\
&&-\int \textrm{d}^3{\bf{x}} \ \frac{\partial}{\partial t}L_{0}^a(x)*\delta^3({\bf{x}}-{\bf{z}})
\label{ku}
\end{eqnarray}
where we have used the property (\ref{b1}). Furthermore, exploiting the cyclicity property of the star product (\ref{b2}), eq. (\ref{ku}) is further simplified so as to bring the Euler derivative at an extreme end, 
\begin{eqnarray}
&&\Lambda^a|_{L_0}({\bf{z}},t)\nonumber\\
&=&-\int \textrm{d}^3{\bf{x}} \ \frac{g}{2}\left(f^{abc}\{\delta^3({\bf{x}}-{\bf{z}}),A^{0c}(x)\}_*+id^{abc}[\delta^3({\bf{x}}-{\bf{z}}),A^{0c}(x)]_*\right)*L_0^b(x)\nonumber\\
&&-\int \textrm{d}^3{\bf{x}} \ \delta^{ab}\delta^3({\bf{x}}-{\bf{z}})*\frac{\partial}{\partial t}L_{0}^b(x).
\label{bdr}
\end{eqnarray}

The same contribution coming from (\ref{eq}) can be written as
\begin{eqnarray}
\Lambda^a|_{L_0}({\bf{z}},t)=\sum_s\int \textrm{d}^3{\bf{x}} \ \frac{\partial^s}{\partial t^s}\left(\rho^{b0 a}_{(s)}(x,z)*L^b_{0}({\bf{x}},t)\right)
\label{sag}
\end{eqnarray}
Only $s=0,1$ contribute so that the above equation simplifies to,
\begin{eqnarray}
\Lambda^a|_{L_0}({\bf{z}},t)=\int \textrm{d}^3{\bf{x}} \ \left(\rho^{b0 a}_{(0)}(x,z)*L^b_{0}({\bf{x}},t)+\rho^{b0 a}_{(1)}(x,z)*\frac{\partial}{\partial t}L^b_{0}({\bf{x}},t)\right).
\label{mo}
\end{eqnarray}
Comparing eqs. (\ref{bdr}) and (\ref{mo}) we obtain,
\begin{eqnarray}
\rho^{b0a}_{(0)}(x,z)&=&-\frac{g}{2}f^{abc}\{\delta^3({\bf{x}}-{\bf{z}}),A_0^c(x)\}_*-i\frac{g}{2}d^{abc}[\delta^3({\bf{x}}-{\bf{z}}),A_0^c(x)]_*
\label{a1}\\
\rho^{b0a}_{(1)}(x,z)&=&-\delta^{ab}\delta^3({\bf{x}}-{\bf{z}}).
\label{a2}
\end{eqnarray}
Other components of the gauge generator can be obtained in a similar way. Here we give the full expressions of these components which will be useful in finding the gauge transformations of the different fields.
\begin{eqnarray}
\rho^{bia}_{(0)}(x,z)&=&-\delta^{ab}\partial^{i{\bf{z}}}\delta^3({\bf{x}}-{\bf{z}})-\nonumber\\
 &&\frac{g}{2}f^{abc}\{\delta^3({\bf{x}}-{\bf{z}}),A^{ic}(x)\}_*-i\frac{g}{2}d^{abc}[\delta^3({\bf{x}}-{\bf{z}}),A^{ic}(x)]_*
\label{37}\\
\phi^a_{i(0)}(x,z)&=&-igT^a_{ij}\delta^3({\bf{x}}-{\bf{z}})*\psi_j(x)\\
\phi'^a_{i(0)}(x,z)&=&-igT^a_{ji}\bar{\psi}_j(x)*\delta^3({\bf{x}}-{\bf{z}})
\end{eqnarray}

Let us next consider the gauge transformations. From eq. (\ref{a}) we write the gauge transformation equation for the zeroth component of the gauge field
\begin{eqnarray}
\delta A^{0 a}({\bf{x}},t)&=&\sum_s(-1)^s\int\textrm{d}^3{\bf{z}} \ \frac{\partial^s\eta^b({\bf{z}},t)}{\partial t^s}*\rho^{a0b}_{(s)}(x,z)\nonumber\\
&=&\int\textrm{d}^3{\bf{z}} \ \left(\eta^b({\bf{z}},t)*\rho^{a0b}_{(0)}(x,z)-\frac{\partial\eta^b({\bf{z}},t)}{\partial t}*\rho^{a0b}_{(1)}(x,z)\right)
\label{mono}
\end{eqnarray}
Exploiting the identity\cite{amorim,rabin}
\begin{eqnarray}
A(x)*\delta(x-z)=\delta(x-z)*A(z)
\end{eqnarray}
 and interchanging $a, \ b$, the generator (\ref{a1}) is recast as,
\begin{eqnarray}
\rho^{a0b}_{(0)}(x,z)&=&\frac{g}{2}f^{abc}\{\delta^3({\bf{x}}-{\bf{z}}),A_0^c(z)\}_*+i\frac{g}{2}d^{abc}[\delta^3({\bf{x}}-{\bf{z}}),A_0^c(z)]_*.
\label{a3}
\end{eqnarray}
Use of the eqs. (\ref{a3}) and (\ref{a2}) along with the identities (\ref{b1}) and (\ref{b2}) in (\ref{mono}) implies that
\begin{eqnarray}
\delta A^{0a}=\partial^{0}\eta^{a}-\frac{g}{2}f^{abc}\{A^{0b},\eta^{c}\}_*+i\frac{g}{2}d^{abc}[A^{0b},\eta^{c}]_*=(\mathcal{D}^{0}*\eta)^a
\label{r1}
\end{eqnarray}
where the operator $\mathcal{D}$ had already been defined in (\ref{D}). In a similar way, using the expression (\ref{37}) we can get the space component of the gauge transformation of the $A^{\mu}$ field as,
\begin{eqnarray}
\delta A^{ia}=\partial^{i}\eta^{a}-\frac{g}{2}f^{abc}\{A^{ib},\eta^{c}\}_*+i\frac{g}{2}d^{abc}[A^{ib},\eta^{c}]_*=(\mathcal{D}^{i}*\eta)^a
\label{r2}
\end{eqnarray}
Combining the two results (\ref{r1}) and (\ref{r2}) we get the following star covariant gauge transformation rule for the gauge field
\begin{eqnarray}
\delta A^{\mu a}=(\mathcal{D}^{\mu}*\eta)^a
\label{chd}
\end{eqnarray}
The same process leads to the star gauge transformation relations of the matter fields as,
\begin{eqnarray}
&&\delta\psi_i(x)=-ig\eta^a(x)*T^a_{ij}\psi_j(x)
\label{cd}\\
&&\delta\bar{\psi}_i(x)=igT^a_{ji}\bar{\psi}_j(x)*\eta^a(x)
\label{cdh}
\end{eqnarray}

Thus the star gauge transformations of all the fields have been systematically obtained. These (eqs. (\ref{chd},\ref{cd},\ref{cdh})) are the results previously stated in Section 2 (eq. (\ref{XX})) under which the action (\ref{lag}) is invariant. Also, the generators $\rho$ are mapped with the gauge identity $\Lambda^a$ ( eq. (\ref{lambda})) by the relation (\ref{lam}). If we set $\theta=0$, then these just correspond to the usual commutative space results for Yang-Mills theory in the presence of matter\cite{gitman}. This implies that, as occurs for the gauge transformations, the mapping (\ref{lam}) is also a star deformation of the usual undeformed (commutative space) map.

Let us now mention a technical point. In obtaining the gauge transformations - say (\ref{r1}) from (\ref{mono})- use is made of identities like (\ref{b1}), (\ref{b2}) which are strictly valid over the whole four dimensional space time. Since (\ref{mono}) involves only the space integral, manipulations based on these identities imply only space-space noncommutativity. This is quite reminiscent of the Hamiltonian formulation of gauge symmetries\cite{rabin} where $\theta^{0i}=0$ from the beginning. 

We conclude this section by providing a simple consistency check. We show that the variation of the action (\ref{lag}) is indeed expressed in the form (\ref{bak}) where $\Lambda^a$ is given by (\ref{lambdaa}). Starting from (\ref{tm}) and using the explicit structures of the variations derived in (\ref{chd},\ref{cd},\ref{cdh}) we obtain,
\begin{eqnarray}
\delta S&=&-\int {\textrm{d}}^4x \ (\mathcal{D}^{\mu}*\eta)^a*L_{\mu}^a+(-ig\eta^a*T^a_{ij}\psi_j)*L_i+(igT^a_{ji}\bar{\psi}_j*\eta^a)*L_i'\nonumber\\
&=&-\int {\textrm{d}}^4x \ \eta^a*\left(\left(-\mathcal{D}^{\mu}*L_{\mu}\right)^a-igT^a_{ij}{\psi}_j*L_i-igT^a_{ji}L_i'*\bar{\psi}_j\right).
\end{eqnarray}
The expression star multiplied with the gauge parameter $\eta^a$ is precisely $\Lambda^a$ given by (\ref{lambdaa}), conforming to the general form (\ref{bak}).
This completes the consistency check.
\section{Analysis for Twisted gauge transformation}
For simplicity we take the pure gauge theory
\begin{eqnarray}
S=-\frac{1}{2}\int \textrm{d}^4x \ {\textrm {Tr}}(F_{\mu\nu}(x)*F^{\mu\nu}(x))
\label{s}
\end{eqnarray}
where the field strength tensor was defined in (\ref{f}). Now the gauge field transforms in the undeformed way
\begin{eqnarray}
\delta_{\eta}A_{\mu}=\partial_{\mu}\eta+ig[A_{\mu},\eta]
\label{gag}
\end{eqnarray}
Using the deformed coproduct rule (\ref{co}) and the gauge transformation (\ref{gag}), the variation of the (star) product of gauge fields is also seen to be undeformed,
\begin{eqnarray}
\delta_{\eta}(A_{\mu}*A_{\nu})=\partial_{\mu}\eta A_{\nu}+A_{\mu}\partial_{\nu}\eta-ig[\eta,(A_{\mu}*A_{\nu})]
\end{eqnarray}
From the above result, gauge transformation of the field strength tensor is now computed
\begin{eqnarray}
\delta_{\eta}F_{\mu\nu}&=&\partial_{\mu}\delta_{\eta}A_{\nu}-\partial_{\nu}\delta_{\eta}A_{\nu}+ig\delta_{\eta}[A_{\mu},A_{\nu}]_*\\
&=&\partial_{\mu}(\partial_{\nu}\eta+ig[A_{\nu},\eta])-\partial_{\nu}(\partial_{\mu}\eta+ig[A_{\mu},\eta])\nonumber\\&&+ig\left([\partial_{\mu}\eta,A_{\nu}]+[A_{\mu},\partial_{\nu}\eta]-ig[\eta,[A_{\mu},A_{\nu}]_*]\right)
\label{GF}\\
&=&-ig[\eta,F_{\mu\nu}]
\end{eqnarray}
Likewise one finds that the expression $F^{\mu\nu}*F_{\mu\nu}$ transforms as,
\begin{eqnarray}
\delta_{\eta}(F^{\mu\nu}*F_{\mu\nu})=-ig[\eta,F^{\mu\nu}*F_{\mu\nu}]
\end{eqnarray}
Both $F_{\mu\nu}$ and $F_{\mu\nu}*F^{\mu\nu}$ have the usual (undeformed) transformation properties. Thus the action (\ref{s}) is invariant under the gauge transformation (\ref{gag}) and the deformed coproduct rule (\ref{co}).

There is another way of interpreting the gauge invariance which makes contact with the gauge identity. Making a gauge variation of the action (\ref{s}) and taking into account the twisted coproduct rule (\ref{co}), we get
\begin{eqnarray}
\delta_{\eta}S&=&-\frac{1}{2}\int \textrm{d}^4x \ {\textrm {Tr}}\delta_{\eta}(F_{\mu\nu}*F^{\mu\nu})\\
&=&-\frac{1}{2}\int \textrm{d}^4x \ [{\textrm {Tr}}(\delta_{\eta}F_{\mu\nu}*F^{\mu\nu}+F_{\mu\nu}*\delta_{\eta}F^{\mu\nu}\nonumber\\&&-\frac{i}{2}\theta^{\mu_1\nu_1}(\delta_{\partial_{\mu_1}\eta}F_{\mu\nu}*\partial_{\nu_1}F^{\mu\nu}+\partial_{\mu_1}F_{\mu\nu}*\delta_{\partial_{\nu_1}\eta}F^{\mu\nu})\nonumber\\&&-\frac{1}{8}\theta^{\mu_1\nu_1}\theta^{\mu_2\nu_2}(\delta_{\partial_{\mu_1}\partial_{\mu_2}\eta}F_{\mu\nu}*\partial_{\nu_1}\partial_{\nu_2}F^{\mu\nu}+\partial_{\mu_1}\partial_{\mu_2}F_{\mu\nu}*\delta_{\partial_{\nu_1}\partial_{\nu_2}\eta}F^{\mu\nu})\nonumber\\&&+\cdot\cdot\cdot)]
\label{delS}
\end{eqnarray}
Now using the result (\ref{GF}) each term of eq. (\ref{delS}) can be computed separately. For example we concentrate on the first term. Using the identity (\ref{b1}) and the trace condition (\ref{trace}) we write the first term as 
\begin{eqnarray}
\delta_{\eta}S|_{{\textrm{1st term}}}&=&-\frac{1}{4}\int \textrm{d}^4x \ (\delta_{\eta}F^{\mu\nu a}*F_{\mu\nu}^a+F^{\mu\nu a}*\delta_{\eta}F_{\mu\nu}^a)\\
&=&-\frac{1}{2}\int \textrm{d}^4x \ \delta_{\eta}F^{\mu\nu a}F_{\mu\nu}^a
\end{eqnarray}
Making use of (\ref{GF}) and dropping the surface terms the above expression is found out to be,
\begin{eqnarray}
\delta_{\eta}S|_{{\textrm{1st term}}}=-\int \textrm{d}^4x&\eta^a&(-\partial^{\mu}\partial^{\nu}F_{\mu\nu}-ig\partial^{\mu}[A^{\nu},F_{\mu\nu}]-ig[A^{\mu},\partial^{\nu}F_{\mu\nu}]\nonumber\\&&+g^2[A^{\mu}*A^{\nu},F_{\mu\nu}])^a
\end{eqnarray}
The second term of (\ref{delS}) is identically zero due to the antisymmetric nature of $\theta^{\mu\nu}$. We write that as,
\begin{eqnarray}
\delta_{\eta}S|_{{\textrm{2nd term}}}&=&-\frac{1}{2}\int \textrm{d}^4x \ \eta^a\frac{i}{2}\theta^{\mu_1\nu_1}(-ig\{\partial_{\mu_1}F^{\mu\nu},\partial_{\nu_1}F_{\mu\nu}\})^a\\
&=&-\int \textrm{d}^4x \ \eta^a\frac{i}{2}\theta^{\mu_1\nu_1}(-ig\{\partial_{\mu_1}\partial^{\mu}A^{\nu},\partial_{\nu_1}F_{\mu\nu}\}\nonumber\\&&+g^2\{\partial_{\mu_1}(A^{\mu}*A^{\nu}),\partial_{\nu_1}F_{\mu\nu}\})^a\\
&=&-\int \textrm{d}^4x \ \eta^a\frac{i}{2}\theta^{\mu_1\nu_1}(-ig\partial^{\mu}\{\partial_{\mu_1}A^{\nu},\partial_{\nu_1}F_{\mu\nu}\}-\nonumber\\&&ig\{\partial_{\mu_1}A^{\mu},\partial_{\nu_1}\partial^{\nu}F_{\mu\nu}\}+g^2\{\partial_{\mu_1}(A^{\mu}*A^{\nu}),\partial_{\nu_1}F_{\mu\nu}\})^a
\end{eqnarray}
The third term is written as
\begin{eqnarray}
\delta_{\eta}S|_{{\textrm{3rd term}}}=-\int \textrm{d}^4x&(\partial^{\mu_1}\partial^{\mu_2}\eta^a)&\frac{1}{2}(\frac{i}{2})^2\theta^{\mu_1\nu_1}\theta^{\mu_2\nu_2}(-\partial^{\mu}\partial^{\nu}\partial^{\nu_1}\partial^{\nu_2}F_{\mu\nu}\nonumber\\&&-ig\partial^{\mu}[A^{\nu},\partial^{\nu_1}\partial^{\nu_2}F_{\mu\nu}]\nonumber\\&&-ig[A^{\mu},\partial^{\nu}\partial^{\nu_1}\partial^{\nu_2}F_{\mu\nu}]\nonumber\\&&+g^2[A^{\mu}*A^{\nu},\partial^{\nu_1}\partial^{\nu_2}F_{\mu\nu}])^a.
\end{eqnarray}
Using the antisymmetry of $\theta^{\mu\nu}$ and dropping the various surface terms, we write the above expression as,
\begin{eqnarray}
\delta_{\eta}S|_{{\textrm{3rd term}}}=-\int \textrm{d}^4x&\eta^a&\frac{1}{2}(\frac{i}{2})^2\theta^{\mu_1\nu_1}\theta^{\mu_2\nu_2}(-ig\partial^{\mu}[\partial^{\mu_1}\partial^{\mu_2}A^{\nu},\partial^{\nu_1}\partial^{\nu_2}F_{\mu\nu}]\nonumber\\&&-ig[\partial^{\mu_1}\partial^{\mu_2}A^{\mu},\partial^{\nu}\partial^{\nu_1}\partial^{\nu_2}F_{\mu\nu}]\nonumber\\&&+g^2[\partial^{\mu_1}\partial^{\mu_2}(A^{\mu}*A^{\nu}),\partial^{\nu_1}\partial^{\nu_2}F_{\mu\nu}])^a.
\end{eqnarray}
Other terms can be obtained in a similar manner. Combining all these terms we finally get,
\begin{eqnarray}
\delta_{\eta}S&=&-\int \textrm{d}^4x \ \eta^a(-\partial^{\mu}\partial^{\nu}F_{\mu\nu}-ig\partial^{\mu}[A^{\nu},F_{\mu\nu}]_*-ig[A^{\mu},\partial^{\nu}F_{\mu\nu}]_*\nonumber\\&&+g^2[A^{\mu}*A^{\nu},F_{\mu\nu}]_*)^a\\
&=&-\int \textrm{d}^4x \ \eta^a\Lambda^a
\end{eqnarray}
where,
\begin{eqnarray}
\Lambda^a=-(\mathcal{D}^{\mu}*L_{\mu})^a=-(\mathcal{D}^{\mu}*\mathcal{D}^{\sigma}*F_{\sigma \mu})^a
\label{70}
\end{eqnarray}
that vanishes identically. Note that this is exactly the same as the expression in the gauge identity (\ref{lambda}) without the Fermionic fields. This proves the invariance of the action.

Let us now repeat the analysis of the previous section with appropriate modifications.
Since the gauge transformations are undeformed, the gauge generators are expected to have the same form
as in the commutative space. To see this note that the gauge variation of the zeroth component of the $A_{\mu}$ field, following from (\ref{gag}), can be written as, 
\begin{eqnarray}
\delta_{\eta}A_0^a(z)&=&\partial_0\eta^a(z)-gf^{abc}A_0^b(z)\eta^c(z)\cr
&=&g\int \textrm{d}^3{\bf{z}} \ f^{abc}A_0^c\eta^b\delta^3({\bf{x}}-{\bf{z}})+\int\textrm{d}^3{\bf{z}} \ \delta^{ab}\delta^3({\bf{x}}-{\bf{z}})\frac{\partial}{\partial t}\eta^b
\end{eqnarray}
Clearly the above result can be expressed in our standard  form (\ref{a}),
\begin{eqnarray}
\delta_{\eta} A_0^a(z)&=&\sum_s(-1)^s\int\textrm{d}^3{\bf{z}}\frac{\partial^s\eta^b({\bf{z}},t)}
{\partial t^s}\rho^{a0b}_{(s)}(x,z)\cr
&=&\int\textrm{d}^3{\bf{z}} \ \eta^b({\bf{z}},t)\rho^{a0b}_{(0)}(x,z)-\int\textrm{d}^3{\bf{z}} \ \frac{\partial \eta^b({\bf{z}},t)}{\partial t}\rho^{a0b}_{(1)}(x,z)
\end{eqnarray}
where
\begin{eqnarray}
&&\rho^{a0b}_{(0)}(x,z)=gf^{abc}A_0^c\delta^3({\bf{x}}-{\bf{z}})
\label{r11}\\
&&\rho^{a0b}_{(1)}(x,z)=-\delta^{ab}\delta^3({\bf{x}}-{\bf{z}})
\label{r22}
\end{eqnarray}
is the gauge generator. Similarly
\begin{eqnarray}
\delta_{\eta} A_i^a(z)&=&\partial_i\eta^a(z)-gf^{abc}A_i^b(z)\eta^c(z)
\end{eqnarray}
is written in the form
\begin{eqnarray}
\delta_{\eta} A_i^a(z)&=&\sum_s(-1)^s\int\textrm{d}^3{\bf{z}} \ \frac{\partial^s\eta^b({\bf{z}},t)}{\partial t^s}\rho^{aib}_{(s)}(x,z)
\end{eqnarray}
for the value
\begin{eqnarray}
\rho^{aib}_{(0)}(x,z)=-\delta^{ab}\partial^{i{\bf{z}}}\delta^3({\bf{x}}-{\bf{z}})+gf^{abc}A_i^c\delta^3({\bf{x}}-{\bf{z}}).
\label{r33}
\end{eqnarray}
No star products appear in the gauge generators $\rho$ and their structure is similar to the undeformed commutative
space expressions. To identify the difference (both from the commutative space results and the star deformed
results) it is essential to look at the gauge identity and its connection with the corresponding gauge generator.

Now as already implied in (\ref{70}), we have a gauge identity for this system, exactly similar to the previous case,
\begin{eqnarray}
\Lambda^a=-\left(\mathcal{D}^{\mu}*L_{\mu}\right)^a=0
\label{su}
\end{eqnarray}
where $L_{\mu}$ is the Euler derivative defined in (\ref{70}). The Euler-Lagrange equation of motion is given by
\begin{eqnarray}
\mathcal{D}^{\sigma}*F_{\sigma\mu}=0
\end{eqnarray}

The gauge identity and the Euler derivatives are mapped by the relation,
\begin{eqnarray}
\Lambda^a({\bf{z}},t)=\sum_{s=0}^n\int \textrm{d}^3{\bf{x}} \ \frac{\partial^s}{\partial t^s}\left(\rho'^{b\mu a}_{(s)}(x,z)L^b_{\mu}({\bf{x}},t)\right)
\end{eqnarray}
where the values of $\rho'^{b\mu a}_{(0)}(x,z)$ and $\rho'^{b\mu a}_{(1)}(x,z)$ are equal to those of $\rho^{b\mu a}_{(0)}$ and $\rho^{b\mu a}_{(1)}$ of the previous example, given in (\ref{a1}), (\ref{a2}) and (\ref{37}). This happens since the Euler derivatives and the gauge identity are identical to those discussed in the previous section. However, here $\rho'$ is not the generator, rather it is $\rho$ (\ref{r11},\ref{r22},\ref{r33}). Consequently $\rho'$  has to be expressed in terms of  $\rho$. To do this, we rewrite eq. (\ref{a1}) under the identification $\rho=\rho'$ as
\begin{eqnarray}
\rho'^{b0a}_{(0)}(x,z)=-\frac{g}{2}f^{abc}\{\delta^3({\bf{x}}-{\bf{z}}),A_0^c(x)\}_*-i\frac{g}{2}d^{abc}[\delta^3({\bf{x}}-{\bf{z}}),A_0^c(x)]_*
\end{eqnarray}
Now making use of the definition of star product (\ref{str}), the above expression is written in the following way
\begin{eqnarray}
\rho'^{b0a}_{(0)}(x,z)&=&-gf^{abc}A^c_0\delta^3({\bf{x}}-{\bf{z}})-g\sum_{n=1}^{\infty}(\frac{i}{2})^n\frac{\theta^{\mu_1\nu_1}\cdot\cdot\cdot\theta^{\mu_n\nu_n}}{n!}\nonumber\\
&&[(\frac{f^{abc}}{2}+i\frac{d^{abc}}{2})\partial_{\mu_1}\cdot\cdot\cdot\partial_{\mu_n}\delta^3({\bf{x}}-{\bf{z}})\partial_{\nu_1}\cdot\cdot\cdot\partial_{\nu_n}A^{0c}(x)\\
&&(+\frac{f^{abc}}{2}-i\frac{d^{abc}}{2})
\partial_{\mu_1}\cdot\cdot\cdot\partial_{\mu_n}A^{0c}(x)\partial_{\nu_1}\cdot\cdot\cdot\partial_{\nu_n}\delta^3({\bf{x}}-{\bf{z}})].\nonumber
\end{eqnarray}
Note that the $\theta$ independent term is nothing but the gauge generator $\rho^{b0a}_{(0)}$ given in eq. (\ref{r11}). Similarly calculating the other components $\rho'^{bia}_{(0)}$ and $\rho'^{b0a}_{(1)}$ from eqs. (\ref{a2}) and (\ref{37}) we obtain,
\begin{eqnarray}
\rho'^{b\mu a}_{(0)}(x,z)&=&\rho^{b\mu a}_{(0)}(x,z)-g\sum_{n=1}^{\infty}(\frac{i}{2})^n\frac{\theta^{\mu_1\nu_1}\cdot\cdot\cdot\theta^{\mu_n\nu_n}}{n!}\nonumber\\
&&[(\frac{f^{abc}}{2}+i\frac{d^{abc}}{2})\partial_{\mu_1}\cdot\cdot\cdot\partial_{\mu_n}\delta^3({\bf{x}}-{\bf{z}})\partial_{\nu_1}\cdot\cdot\cdot\partial_{\nu_n}A^{\mu c}(x)
\label{tui}\\
&&(+\frac{f^{abc}}{2}-i\frac{d^{abc}}{2})
\partial_{\mu_1}\cdot\cdot\cdot\partial_{\mu_n}A^{\mu c}(x)\partial_{\nu_1}\cdot\cdot\cdot\partial_{\nu_n}\delta^3({\bf{x}}-{\bf{z}})]\nonumber\\
\rho'^{b0a}_{(1)}(x,z)&=&\rho^{b0a}_{(1)}(x,z).
\end{eqnarray}

We conclude that although the generator remains undeformed, the relation mapping the gauge identity with the generator is neither the commutative space result nor its star deformation as found in the other approach. Rather, it is twisted from the undeformed result. The additional twisted terms are explicitly given in (\ref{tui}). Also, the structure of the generator shows that all gauge groups, instead of just $U(N)$ as happened for star gauge transformations, are allowed
\section{Conclusions}
Gauge symmetries on canonically deformed coordinate spaces were considered. Both possibilities (namely, deformed gauge transformations keeping the standard Leibniz rule intact or undeformed gauge transformations with a twisted Leibniz rule) were analysed within a common framework. Explicit structures of the gauge generators were obtained in either case. The connection of these  generators with the gauge identity, which must exist whenever there is a gauge symmetry, was also established. In the former case, this connection was a star deformation of the commutative space result. In the latter case, on the other hand, the commutative space result was appropriately twisted. It is quite remarkable that these fundamental properties of gauge symmetries (i. e. occurrence of gauge identity and its connection with the corresponding generator through the 
Euler derivatives) were found in the noncommutative theory, adopting either of the two interpretations. This suggests that deformed gauge theories have properties similar to what we desire for physics, at least as far as gauge symmetries are concerned. All results obtained here reduce to the usual commutative space expressions in the limit of vanishing
$\theta$.

\end{document}